# Robust excitation and Raman conversion of guided vortices in chiral gas-filled photonic crystal fiber

S. DAVTYAN*, Y. CHEN, M. H. FROSZ, P. ST.J. RUSSELL AND D. NOVOA

*Max-Planck Institute for the Science of Light, Staudtstrasse 2, 91058 Erlangen, Germany*
*Corresponding author: sona.davtyan@mpl.mpg.de*

**The unique ring-shaped intensity patterns and helical phase fronts of optical vortices make them useful in many applications. Here we report for the first time efficient Raman frequency conversion between vortex modes in twisted hydrogen-filled single-ring hollow core photonic crystal fiber (SR-PCF). High fidelity transmission of optical vortices in untwisted SR-PCF becomes more and more difficult as the orbital angular momentum (OAM) order increases, due to scattering at structural imperfections in the fiber microstructure. In helically twisted SR-PCF, however, the degeneracy between left- and right-handed versions of the same mode is lifted, with the result that they are topologically protected from such scattering. With launch efficiencies of ~75%, a high damage threshold and broadband guidance, these fibers are ideal for performing nonlinear experiments that require the polarization state and azimuthal order of the interacting modes to be preserved over long distances. Vortex coherence waves (VCW) of internal molecular motion, carrying angular momentum, are excited in the gas, permitting the polarization and orbital angular momentum of the Raman bands to be tailored, even in spectral regions where conventional solid-core waveguides are opaque or susceptible to optical damage.**

---

The study of optical vortices —laser beams carrying orbital angular momentum (OAM) [1] —has led to applications in optical micromanipulation [2], imaging [3], super-resolution microscopy [4], telecommunications [5] and astrophysics [6], as well as encryption of information in both quantum [7] and classical [8] systems. An optical vortex is characterized by the presence of one or more phase singularities where the field intensity vanishes. The phase advance around the singularity, measured by passing the beam through a linear polarizer and interfering it with a phase-scanned linearly polarized reference beam, is an integer multiple $\ell$ of $2\pi$, where $\ell$ is the OAM order (or topological charge).

Long-distance transport of optical vortices is difficult because their phase structure is sensitive to external perturbations, making long free-space beamlines difficult to implement. Special silica-based solid-core fibers partly resolve this problem [9], although at the expense of strong dispersion, low damage threshold and limited transmission bandwidth, which impairs the robust propagation of vortex pulses in difficult-to-access spectral regions such as the ultraviolet and the mid-infrared. In addition, although non-destructive manipulation of the spectral content of optical vortices is of great importance in many of the research fields mentioned above, it is not easily achievable. Most common solutions to frequency conversion of OAM-carrying beams exploit four-wave mixing in free-space arrangements [10,11], stimulated Raman scattering (SRS) in crystals [12] or high-harmonic generation [13]. Only very recently, a scheme using fibers with a ring-shaped silica core was reported [14].

In this Letter we report for the first time Raman frequency conversion between optical vortex modes in hydrogen-filled twisted single-ring photonic crystal fiber (SR-PCF). Being birefringent in both spin and orbital angular momentum (OAM), twisted SR-PCF offers topological protection to vortex modes. High launch efficiencies (reaching 75% in the experiments), low attenuation over a wide spectral bandwidth and tight modal confinement render the gas-Raman interactions highly efficient [15].

In SRS, pump light is non-elastically scattered by Raman-active molecules, generating a red-shifted Stokes signal that beats with the pump light, creating a coherence wave (CW) of synchronized molecular motion that further amplifies the Stokes signal. Within its coherence lifetime, this CW can be used to frequency up-shift light, provided phase-matching is satisfied [16]. It was previously shown that the circular birefringence of the $LP_{01}$-like mode in helically-twisted SR-PCF permits precise control of the polarization state of pump, Stokes and anti-Stokes light in rotational SRS [17].

Here we extend this work to the case when the signals and the CWs carry vortices, enabling transfer of OAM between the Raman sidebands [18]. Although vortex modes in twisted SR-PCF are to a very good approximation circularly polarized, this breaks down close to the core edge where the capillaries protrude into the modal field. It turns out that the azimuthal order $\ell_A$, the number of complete periods of phase progression around the azimuth for fields evaluated in cylindrical coordinates, is a robustly conserved quantity in all $N$-fold rotationally symmetric PCFs, no matter what their structure [19]. In addition, although for circularly polarized fields the OAM order is given by $\ell = \ell_A + s$ ($s = -1$ for right-circular polarization, RCP), for more complex field distributions, such as seen in SR-PCF, this is no longer true since $|s| \neq 1$. In contrast to circularly symmetric systems, the $N$-fold rotational symmetry of the fields also requires the presence of azimuthal harmonics

$\ell_A^{(m)} = \ell_A + mN$ where $m$ is an integer and $\ell_A$ lies within the first Brillouin zone, i.e., $-N/2 < \ell_A < N/2$ [20]. Deviations from perfect $N$-fold symmetry in untwisted SR-PCF cause unpredictable amounts of linear birefringence that disturb the spin and OAM of a launched vortex mode [21]. The circular birefringence in twisted SR-PCF suppresses these effects, permitting undistorted transmission of optical vortex modes (Fig. 1a).

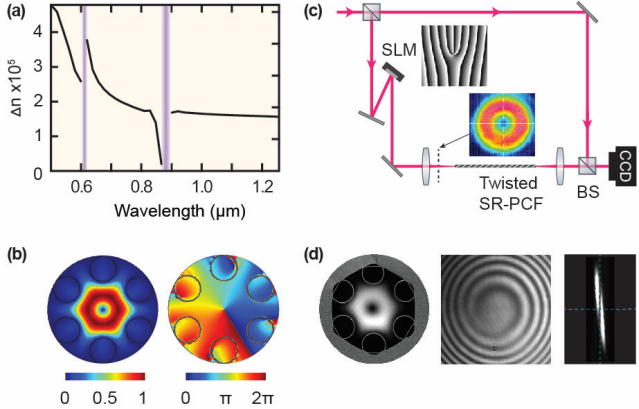

Fig. 1. (a) Index difference Δn between two twisted SR-PCF modes with azimuthal order $\ell_A = 0$ and $s = \pm 1$ ($\ell = \pm 1$), calculated by finite-element modelling (FEM). The shaded purple regions indicate spectral regions of high loss caused by anti-crossings with resonances in the capillary walls. (b) Calculated intensity and phase patterns of the $\ell_A = 0$, $s = -1$ mode at 1030 nm in a twisted SR-PCF with perfect 6-fold symmetry. (c) Experimental set-up for excitation and characterization of vortex modes. The insets show the SLM phase pattern and the measured transverse field at the dashed plane. (d) Left: Measured near-field intensity pattern at the output of a 2-m long twisted SR-PCF, overlaid on a scanning electron micrograph of the microstructure. Middle: Interference pattern generated by superimposing the vortex mode and a spherical reference wave. Right: Image cast on a screen by the vortex mode, after passing through a cylindrical lens.

Figure 1b shows the experimental set-up used to generate and characterize vortex modes in twisted SR-PCF. A pulsed 1030 nm TEM$_{00}$ laser beam is first divided at a beam splitter. The first part is sent to a reflective phase-only SLM with a forked phase mask imprinted on it (see inset in Fig. 1c), resulting in production (in the first diffracted order) of an $\ell = -1$ Laguerre-Gaussian (LG) beam with a phase singularity at its center and a spiraling phase front ($s = -1$). This beam was launched into a 2-m length of twisted SR-PCF (twist rate 0.5 rad/mm) with 50 μm core diameter, ~22 μm capillary diameter and mean capillary wall-thickness 820 nm. By keeping the fibers straight so as to minimize bend-losses, we were able to increase the transmission up to 65%, corresponding to a launch efficiency of ~75%, assuming 1 dB/m attenuation. The twist was introduced by spinning the preform during fiber drawing. Apart from narrow bands of high loss caused by anti-crossings between the core mode and resonances in the capillary walls, the fiber provided broadband guidance with a loss of ~1 dB/m.

A clear six-fold symmetric vortex mode emerges from the fiber end-face (Fig. 1d). Its helical phase was confirmed by interference with a divergent beam formed from the second pump beam, revealing the spiral interference pattern characteristic of an $\ell = -1$ singularity, identical to the launched beam. The OAM order of the Raman bands (for which there was no convenient reference beam) was measured using a cylindrical lens (Fig. 1d), which transforms the incoming LG beam into a set of Hermite-Gaussian modes [22]. In spite of imperfect overlap between the launched LG beam and the vortex mode in the SR-PCF, more than 50% of pump power emerged in a vortex beam from the end-face of the twisted SR-PCF. Carrying out the same experiment in an untwisted SR-PCF with a closely similar microstructure showed very poor preservation of the OAM order, highlighting the key role played by the twist.

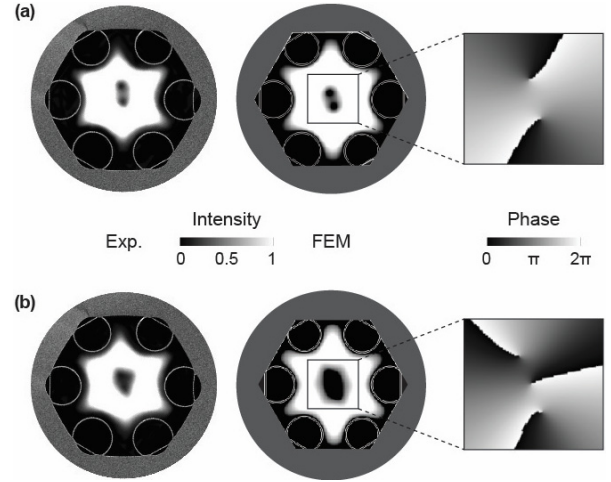

Fig. 2. (a) Left: Measured field intensity pattern at the output of a 2-m-long twisted SR-PCF for an $\ell = +2$ ($\ell_A = +3$ and $s = -1$) mode, saturated so as to highlight the presence of two spatially separated zero-field singularities at core center [23]. Middle: FEM simulations of the field intensity based on a structure with distortions similar to those in the actual fiber. The agreement with experiment is good. Right: Zoom into the phase distribution close to core center, showing the phase singularities. (b) Same as (a), but for a mode with $\ell = +3$ ($\ell_A = +4$ and $s = -1$).

As mentioned above, OAM modes are in general susceptible to perturbation by structural inhomogeneities, with a sensitivity that increases with OAM order. The consequence of this may be seen in near-field images of the intensity pattern at the output of the SR-PCF when modes with OAM order $\ell = +2$ (Fig. 2a) and +3 (Fig. 2b) are launched. Although the total OAM order is topologically preserved, structural imperfections in the fiber microstructure cause the topological charges to separate out into two or three off-axis $\ell = +1$ phase singularities, as predicted by Nye and Berry [23]. The experimental results agree well with FEM modeling, based on scanning electron micrographs of the actual SR-PCF microstructure.

The rotational Raman gain (frequency shift 17.6 THz in hydrogen) is higher for circularly polarized pump light, and if axial linear momentum and spin are appropriately conserved, efficient generation of anti-Stokes light is also possible [17]. The experimental set-up for investigating rotational Raman frequency conversion of OAM modes in SR-PCF is sketched in Fig. 3. Part of the pump power is used to generate a seed signal at 1097 nm (the first rotational Stokes band) [24] in a photonic band-gap fiber (PBG) filled with 20 bar of $H_2$. The rest of pump light is converted from TEM$_{00}$ to an LG mode carrying OAM (see Fig. 1). The polarization states of pump and seed are adjusted with a combination of quarter-wave plate and polarizing beam-splitter (PBS). The path-lengths are adjusted so that the pump and seed pulses overlap at a dichroic mirror (DM), where they are combined and launched into a 70 cm length of twisted SR-PCF filled with 25 bar of $H_2$.

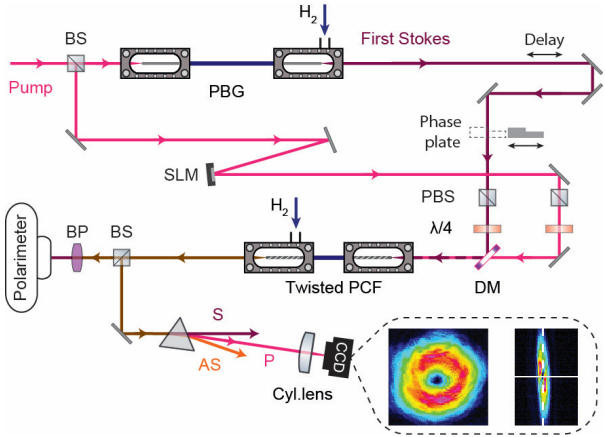

Fig. 3. Set-up for studying seeded SRS with vortex modes. PBG is a photonic bandgap PCF with a passband that guides only the pump (1030 nm) and first rotational (1097 nm) Stokes signals. (Inset) Images of the pump intensity profile ($\ell_A = 0$ and $s = -1$) recorded with a CCD camera: collimated (left) and focused with a cylindrical lens (right).

A portion of the output light is split off at a beam splitter and the spin of each Raman sideband measured using a set of band-pass filters (BP) and a home-built polarimeter. The remaining portion is sent through a dispersive $CaF_2$ prism to separate out its spectral components, and the OAM state of each comb-line separately measured. For a pump with $\ell_A = 0$ and $s = -1$ ($\ell = -1$) and a Stokes seed with $\ell_A = -1$ and $s = +1$ ($\ell = 0$), we observed amplification of the seed and the emergence of an anti-Stokes signal with $\ell_A = -1$ and $s = -1$ ($\ell = -2$). The far-field intensity profiles of these modes are shown in the right-hand column in Fig. 4a. They were recorded 5 meters away from the fiber end-face so as to ensure sufficient spatial separation of the sidebands (the anti-Stokes profile is distorted because of the topological charge splitting discussed above).

To interpret these results, we make use of generalized dispersion diagrams (Fig. 4a) in which the Raman transitions are plotted as a function of azimuthal order within the first Brillouin zone $-3 < \ell_A < +3$. The blue arrows represent the change in azimuthal order and spin in a transition between a pump with $\ell_A = 0$ and $s = -1$ and a Stokes seed with $\ell_A = -1$ and $s = +1$. The VCW reverses the spin while providing the necessary change in linear momentum $\Delta \beta = \beta_P - \beta_S$ and azimuthal order $\Delta \ell_A = \ell_A^P - \ell_A^S = +1$ where P = pump and S = Stokes. Note that a VCW created in this way can only be used, within its lifetime, to frequency upshift light from the pump to the anti-Stokes if a second "mixing" pump signal with the opposite spin is launched (hollow circles in Fig. 4a) where the two upper blue arrows show transitions from $\ell_A^P = 0$ to $\ell_A^{AS} = +1$ and $\ell_A^P = -2$ to $\ell_A^{AS} = -1$, as confirmed by the experiments. This can occur if the pump is slightly elliptically polarized. The dashed grey arrows in Fig. 4a illustrate the case when the azimuthal order and spin of pump and seed are reversed, resulting in generation of an anti-Stokes band at $\ell_A^{AS} = -1$. The azimuthal order of the $n$-th order upshifted anti-Stokes signal generated from a mixing signal (M) at arbitrary frequency is $\ell_A^n = \ell_A^M + n\Delta\ell_A$ [12].

To further test the selection rules, we studied a second configuration involving copolarized $s = -1$ pump and Stokes modes, both with $\ell_A = 0$ (Fig. 4b). The Stokes OAM ($\ell = -1$) was introduced using a commercial phase plate. It is remarkable that, by seeding with a co-polarized Stokes, the much higher gain spin $-1$ to $+1$ transition is suppressed by the polarization-maintaining property of the twisted SR-PCF (see also [17]), resulting in an anti-Stokes signal that appears in a clear vortex mode with the same OAM and spin as the input beams.

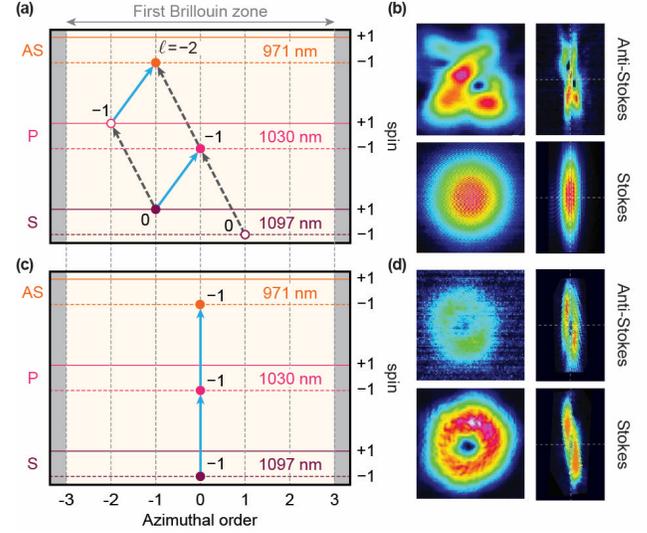

Fig. 4. (a)&(c) State diagrams for Stokes and anti-Stokes transitions in two different cases. The horizontal axis marks the azimuthal order $\ell_A$ and the full and dashed horizontal lines correspond to $s = +1$ and $-1$ respectively. (a) The blue arrows mark the allowed transitions for a VCW excited by an $\ell_A = 0, s = -1$ pump and an $\ell_A = -1, s = +1$ Stokes seed (OAM orders $\ell$ marked in). The dashed arrows are for an $\ell_A = 0, s = -1$ pump and an $\ell_A = +1, s = -1$ Stokes seed. (b) Mode intensity profiles measured after collimation (left) and field patterns close to the focal plane of a cylindrical lens (right). (c) Transitions for the case when pump and Stokes seed both have $\ell_A = 0, s = -1$. The spin $-1$ to $+1$ transition, which has higher gain, is suppressed by the twisted SR-PCF. (d) Intensity profiles and field patterns for (c).

To illustrate that hydrogen-filled twisted SR-PCF can allow efficient generation of vortex modes in spectral regions far away from the pump frequency, we generated a strong vibrational Stokes signal at 1.8 μm using linearly-polarized pump pulses formed by a superposition of $\ell_A = 0, s = -1$ and $\ell_A = -2, s = +1$ modes. Note that although these two modes are in principle non-degenerate, the beat length is ~10 m (for a circular birefringence of ~$10^{-7}$), i.e., much longer than the fiber. As a result there is negligible optical rotation along the fiber, and the pump polarization state is transferred cleanly to the Stokes signal. Focusing through a cylindrical lens we found that the signal at 1.8 μm had OAM order $\ell = +1$ (inset of Fig. 5a), meaning that it was formed by the superposition of $\ell_A = 2, s = -1$ and $\ell_A = 0, s = +1$ modes. This was unexpected, since the nonlinear gain is supposed to be similar for the $\ell = -1$ Stokes mode [14]. We attribute the weaker $\ell = -1$ Stokes signal to higher fiber loss, as predicted by FEM. The weak rotational anti-Stokes sideband at 1.63 μm appears as a result of modulation of the 1.8 μm signal by the rotational VCWs generated by the pump and its rotational Stokes at 1097 nm.

We tested the quantum conversion efficiency at 25 bar hydrogen by measuring the output photon rates for increasing pump energy

(Fig. 5b). A maximum of ∼30% was measured at a pump energy of ∼100 μJ. Good agreement with numerical solutions of Maxwell-Bloch equations [25] was found when the loss of the vibrational Stokes band was set to 8 dB/m (solid lines). It is caused by an anti-crossing with resonances in the capillary walls. A twisted SR-PCF with thinner capillary walls would move this resonance to shorter wavelengths, increasing the efficiency. Numerical simulations predict a quantum efficiency of 50%, assuming pump and Stokes losses of ∼1 dB/m (orange dashed line).

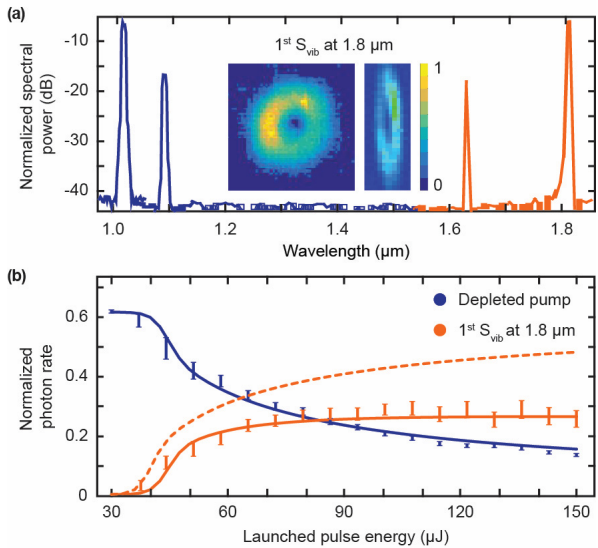

Fig. 5. (a) Uncalibrated spectra for ∼83 μJ of pump energy measured with an optical spectrum analyzer (blue) and an InGaAs spectrometer (orange). All the lines are normalized to the peak pump intensity. (Inset) Image of the 1st vibrational Stokes signal at 1.8 μm collimated (left) and focused by a cylindrical lens (right). (b) Photon flux of the depleted pump (blue) and the 1.8 μm Stokes (orange) signals normalized to the pump flux before entering the fiber (60% launch efficiency). Simulations (solid curves) agree well with the measurements. The dashed orange curve plots the theoretical Stokes conversion efficiency when the fiber attenuation is negligible (see text).

In conclusion, gas-filled twisted SR-PCF provides a promising platform for precise excitation, robust polarization-preservation and efficient Raman frequency conversion of optical vortices, at the same time offering low-loss guidance. As a result, the polarization state and OAM of SRS sidebands can be widely tailored, making possible efficient broadband frequency conversion between vortices in difficult-to-access spectral regions where conventional approaches fail. The use of twisted versions of recent SR-PCF designs offering record low loss [26] will allow even higher SRS quantum efficiencies. We believe the results reported here will lead to a revolution in guided singular optics, with applications across many fields of science.

**Acknowledgements.** We thank Andrea Cavanna and Sébastien Loranger for help with several aspects of the experiments.